# This is the author version of the published paper

# A Discrete Event System Specification (DEVS)-Based Model of Consanguinity

Noreen Akhtar, Muaz Niazi[1, 2]*, Amir Hussain[2] and Farah Mustafa[1]

[1] Department of Biosciences,

COMSATS Institute of IT,

Islamabad, Pakistan

[2] Department of Computing Science and Mathematics,

University of Stirling,

Scotland, UK

**Abstract**

*Consanguinity or inter-cousin marriage is a phenomenon quite prevalent in certain regions around the globe. Consanguineous parents have a higher risk of having offspring with congenital disorders. It is difficult to model large scale consanguineous parental populations because of disparate cultural issues unique to regions and cultures across the globe. Although consanguinity, as a social problem has been studied previously, consanguinity from a biological perspective has yet to be modeled. Discrete Event System Specification (DEVS) formalism is a powerful modeling formalism for the study of intricate details of real-world complex systems. In this article, we develop a DEVS model to get an insight into the role of consanguineous marriages in the evolution of congenital disorders in a population. As proof-of-concept, we develop a consanguinity simulation model in Simio simulation software. Our results show the effectiveness of DEVS in the modeling of consanguinity effects in causing congenital defects.*

# 1. **Introduction**

A consanguineous or inter-cousin marriage is a cultural tradition in many societies around the world [1]. Consanguineous marriage is formally defined as "a marriage which is solemnized among persons descending from the same stock or common ancestor with close biological relations" [2]. Although consanguinity can have positive effects such as increase in general population fitness and a reduction in breast cancer [3, 4], at the same time, it has also been known to lead to an increased rate of birth defects, manifesting as severe recessive disorders [5-8]. Various studies have pointed out that consanguinity, a cultural trait, at times lowers certain population fitness factors [9-11]. Despite this, inter-cousin marriages are prevailing and in fact spreading because of their socioeconomic usefulness amongst diverse populations.

Outside its social and cultural context, consanguinity from a biological perspective has not been modeled in the past. Computational Modeling and simulation techniques have previously proven useful in developing insights and understanding of the dynamics of complex biological systems [12, 13]. Large scale consanguineous parental population is in essence, a possible domain for the application of simulation. However, its emergence in societies within the same geographic area, despite cultural, linguistic and religious heterogeneity [14, 15] make its modeling a challenging problem. Discrete Event System Specification (DEVS) formalism [16], a formal modeling and simulation framework, has been successfully used as a framework for modeling complex systems. Although DEVS has been used to model certain natural and biological systems [17, 18], it has not been used to model consanguinity to the best of our knowledge.

The purpose of the present study was to examine the role of consanguineous marriages in causing congenital defects using a modeling and simulation approach. With a dearth of modeling and simulation studies in the domain of consanguinity, there is need to develop formalism for modeling this complex system. In this paper, we focus on the development of a DEVS framework for the formal modeling of consanguinity. As a proof-of-concept, we further demonstrate conversion of this model to an actual simulation model using Simio simulation software [19]. Our results show that DEVS can be used effectively to model biological problems.

## 2. Background

The widespread practice of consanguineous marriages has conventionally been attributed to its multiple social benefits, such as aggregation of economic wealth, better treatment of spouse, increased family stability and security [20, 21]. However, it has also been well demonstrated that consanguineous marriages have a relatively higher risk of producing offsprings with genetic disorders than that of the general population [22, 23]. These include diabetes mellitus, cancers such as that of the cervical, brain, etc. and coronary artery disease as discussed in several articles [8, 11, 24, 25]. Consanguinity has even been considered to contribute to an increase in incidents of hypertension [26, 27]. The detrimental health effects associated with consanguinity are caused by the expression of recessive genes inherited from a common ancestor(s) [2]. This applies to rare single gene conditions as well as to multigene disorders with multifactorial inheritance. Therefore, it is often proposed that consanguineous marriages should be discouraged on medical grounds.

The degree of relationship involved in consanguineous marriages affects the rate of birth defects proportionally. Three relationship degrees are considered to have deleterious effects on human health that include first, second and third degree cousins as shown in Figure 1. First-cousin marriages are the most common type of consanguineous union because they share twice the consanguinity (four times the degree of consanguinity of second cousins) as any other degree relationship and are used as prototypical examples in studies of consanguinity [14]. On the other hand, first cousins once removed have half the shared DNA as full first cousins. Sometimes, even half-fourth cousins cannot be detected at the DNA level [28].

Due to the complexity of the population interactions regarding consanguinity and large number of components involved, it is difficult to effectively study the behavior of consanguineous population along with congenital disorders. Many statistical studies have been conducted to study these interactions, but these studies have provided limited information regarding the resultant effects of consanguinity on a given population [15, 29, 30]. A few studies regarding simulation of consanguinity networks exist [31, 32], but they have treated consanguinity as a social problem rather than a biological one. Since modeling and computer simulation techniques have previously proven useful for developing an understanding of the dynamics of complex biological systems [33, 34], therefore this approach has been used to study consanguinity here.

## 2.1 Modeling and Simulation

Modeling is the process of producing models for a simulation study [35]. The main focus of modeling is on the input and output signal relation instead of detailed dynamics within the system [36]. Simulation is a tool which is used to simulate an abstract model or generate behavior of a particular system. Simulations are implemented with the help of simulators. If a model is a set of mathematical instructions, then simulator is a software which is used to execute these instructions and generate the behavior of the system of interest [37]. The framework of modeling and simulation consists of four main entities [38] :

- Experimental frame
- Real/virtual source system to be simulated
- The model
- The simulator

Experimental frame specifies environment or conditions under which the system is experimented with. The source system is the real or virtual environment that is to be modeled and data is gathered by observing it. The model is a mathematical representation of a system or structure for generating behavior claimed to represent the real world. The simulator is that computational system/software which obeys instructions of the model and generates behavior shown in Figure 2 [38].

The modeling and simulation entities become significant only when they are properly related to each other. "Modeling Relation" is concerned with how well the "Model" generates behavior and agrees with observed system behavior, while "Simulation Relation" ensures that the simulator carries the model instructions correctly [38]. The framework of modeling includes many system specification formalisms, such as differential equation system specification (DSS), discrete time system specification (DTSS), and discrete event system specification (DEVS). These formalisms help to model systems in the most appropriate and effective manner during the early development at the requirements and specification levels [38].

The modeling and simulation entities become significant only when they are properly related to each other. "Modeling Relation" is concerned with how well the "Model" generates behavior and agrees with observed system behavior, while "Simulation Relation" ensures that the simulator carries the model instructions correctly [38]. The framework of modeling includes many system specification formalisms, such as differential equation system specification (DSS), discrete time system specification (DTSS), and discrete event system specification (DEVS). These formalisms help to model systems in the most appropriate and effective manner during the early development at the requirements and specification levels [38].

Modeling and computer simulation techniques have previously proven useful for understanding the dynamics of complex biological systems [12]. Recently, DEVS formalism has been used as a framework to model natural or biological systems effectively, such as [18, 39, 40]. Therefore, in this study, we proposed use of the DEVS formalism to model the potential effects of consanguinity in causing congenital defects.

### 2.1.1 Discrete Event System Specification

DEVS is a formal mathematical framework which is used to design models for discrete event simulation [37]. DEVS models are usually described as either atomic or coupled models which are defined as tuples:

$$(X, Y, S, \delta ext, \delta int, \lambda, ta) \quad \text{(Atomic model)}$$

$$(X, Y, D, \{M_i\}, \{I_i\}, \{Z_{i,j}\}, select) \quad \text{(Coupled model)}$$

Usually these DEVS models offer specifications like inputs (X), outputs (Y), set of states (S), time advance (ta) and functions for determining next states and outputs given current states and inputs, etc. The major scheme behind DEVS is that the model and simulator work separately and the simulator does not depend on model in a sense that it can run simulations regardless of what DEVS model represents [18].

DEVS formalisms are developed to improve system reliability, design time and comprehensibility. Therefore, DEVS formalism provides a good framework to model consanguinity as risk factor for many congenital disorders because it is a mathematical paradigm with well-defined concepts of coupling of components, hierarchical, modular construction, and support for discrete events [41]. The impact of consanguinity as a genetic risk factor is not modeled yet using formal methods. Therefore, our aim was to provide a DEVS framework to model consanguinity followed by conversion of DEVS model in simulation using Simio software.

**2.1.1.1 DEVS in relation to other approaches:**

For the specification of model frameworks various formalisms have been in use from decade. For example Petri Nets, Discrete Time Specification, Finite state Automata and Queuing networks has been widely used to specify system properties. Choice of any of these formalisms is entirely based on the application domains, modeler's background, goals, or the available computational resources. In case of Discrete Time Specification formalism inconsistency in the system states specification is the major drawback because it treats time variable as a **constant** number whereas in contrast a real world system continuously change with time. DEVS formal model specification framework, which uses mathematical notations to specify a system's behavioral characteristics, overcomes this problem by mainly focusing on the time ''t'' variable, whose value continuously changes like other variables in a system. It do so by separating the system states and constant states by using transition functions which calculate constant states from current system states. DEVS separately keep the simulators and models to handle system states and constant states complexity.

For the modeling of complex biological systems various statistical and computational methods are in use but the development of complex models that can be used to simulate data is not always intuitive. Statistical distributions provide estimations in the form of numbers and percentages therefore these are less reliable in terms that they cannot handle heterogeneities and micro behaviors. Likewise many algebraic notations do not include the time element that DEVS inherits from its system theoretic origins. Genetic Algorithm (GA), one of the computational methods, must requires a solution for the problem generated by the developers therefore cannot be implemented where effects of biological systems are complex, nonlinear, and partially or solely dependent on the effects of other factors. DEVS is well adapted to be implemented in an object oriented framework, thus creating a component based modeling and simulation environment to model complex biological system where one component function depends on the other component which has compelled us to investigate its application to model consanguinity as complex biological system.

In recent most years DEVS become quite supportive in specifying complex biological systems such as biological cell behaviors in cell networks and to capture the motion of deformable biological structures (specified in section 1). The reason which motivates researchers to use DEVS framework is that it offers a full range of computational means such as it provide an ease to the modeler to specify models directly in its terms and system managability when disintegrated model and simulators are required. In our study we used DEVS and SIMIO simulation software for modeling and simulation purpose. So in our opinion DEVS formalism is relatively useful to model consanguinity as it provides facility to handle complexity.

## 3. Development of the DEVS Model of Consanguinity

This section describes the DEVS model of consanguinity as a risk factor. As mentioned earlier, framework development of consanguinity system identifies the key elements or entities and their relationships. Therefore, first we specified the basic entities of modeling and simulation of consanguinity, i.e., source/real system along with the experimental frame, model and simulator. The experimental frame included region, religion, arranged marriage or self commitment, % consanguinity, allelic frequency, consanguinity type (i.e., $1^{st}$ or $2^{nd}$ degree), and co-efficient of inbreeding. With consanguinity as the source system, the DEVS model and Simio simulator are shown in Figure 3.

After specifying the entities, the DEVS formalism of consanguinity was built which is described below:

**M = (X, S, Y, δint, δext, λ, ta)** where

    **M** is model representing the overall system

    **X: set of input values** i.e. (Region, religion, arrange/self commitment, % consanguinity, allelic frequency, consanguinity type, and co-efficient of inbreeding)

    **S: set of states** i.e. Birth Event → Marriage Event → Birth Event → Marriage Event

(Between these two events, the probability of congenital disorders and deaths are also included)

**Y: set of output values, i.e**., new populations with probability of congenital disorders due to consanguineous and non-consanguineous marriages

**δint: Internal transition function,** i.e., if birth event occurs in consanguineous union, then there is a probability of congenital disorder risk and death events.

**δext: External transition function,** i.e., if male and female populations are available, then consanguineous marriage event will occur.

**λ: Output Function,** i.e., S→Y

**ta: Time advance function,** i.e., new births, deaths and marriages per year

The model development begins with a population growth submodel in order to understand how marriage and population growth events occurr, and then focus on DEVS-based design of consanguinity to demonstrate its complexity. In the population growth model (Figure 4), initially two populations are used, i.e., male (MP) and female (FP) rather than whole population to simplify the model development and analysis. After separating both populations, marriage events occur followed by population growth. Under population growth, another source of offspring is added to generate children from resulting marriages randomly. This leads to a new population consisting of the number of total marriages and offspring. To simplify its presentation, we focused on the handling of interactions between male and female population.

When designing a particular DEVS-based system model, one must address two fairly distinct sub-problems: interaction detection and interaction response of male and female populations. Recognizing this, one can simplify matters by focusing on detection and response separately. In our model, we made this separation explicit by defining the DEVS as a coupled model.

The population growth submodel design developed, helped us to expand the model of consanguinity as a risk factor. Then DEVS-based consanguinity formal model was developed with embedded population growth submodel (Figure 5), which has been used to simulate certain interactions in human population which shows the formalism's potential as a means of addressing the complexity of spatial biological models. In this model, initially the whole population (WP) is used and then separated into two groups, i.e., male (MP) and female (FP). This allowed us to obtain male and female entities equally for marriage events. After combining both male and female populations, is the group was split into two, i.e. consanguineous and non-consanguineous marriages. Consanguineous and non-consanguineous marriages event now occurred separately followed by population growth (births). At the end, we obtained a new population with the probability of consanguineous and non-consanguineous population with their rate of offspring growths.

Since the usefulness of a DEVS-based model design depends largely on the extent to which an implementation adheres to it; we therefore choose a simulation approach to implement the DEVS-based consanguinity model.

# 4. Simulation

In the previous sections, we presented basic entities and framework of consanguinity using DEVS formalism. This section will further describe the implementations of consanguinity using the simulation software Simio. Simio makes simulation easier for decision making and enables users to solve more problems, more easily than ever before. It is based on the model object-oriented framework and facilitates building of 3D models [48]. In Simio the basic concept of object oriented framework is that classes define the behavior of objects [19]. Those classes, when placed together in a model, result in the emergence of system behavior from previously defined object interactions. Objects can be user defined and can easily be added and extended in Simio. The basic object types in Simio are [48]:

- Fixed (a fixed location)

- Source (generate entity objects)

- Server (model a capacitated process)

- Sink (destroy entities that have finished processing in the model)

- Link (paths between objects)

- Node (intersection between links)

- Agent (unconstrained movement through free-space)

- Entity (agent) it moves across links, enter objects

- Transporter (entity) it carries entities

- Combiner (attaches the batched members to a parent entity)

- Separator (separates batched members to a parent entity or makes copies of entity objects)

Model development in Simio begins with a population growth submodel, based on the flow diagram depicted in Figure 4. Many objects are used in this simulation, such as model entities (here entities mean individuals or parents), source, server, combiner, paths and sink which destroy entities. Two source objects are used to create the male and female populations to facilitate the correct manner of handling male and female entities. Then both of the source populations are combined to a 'Combiner' object that takes one of each entity, combines them (marriage), and sends them to Server1 named "Population Growth". Connecting the female-population source to the top entry point of the Combiner and connecting the male-population source to the bottom entry point of the combiner takes a parent and member. In this case, the default batch quantity of '1' is used since we require only one male and one female entity to attach.

To execute "Population Growth" event, a server is used within the "Add-On Process Triggers" section of "Properties", which adds a new logic named a 'Processed'. This directs the user to the "Processes" window with a new process called "Server1_Processed". Within this process, we can use a 'Create' step to create new entities (offspring) based on a distribution to determine the number of objects to create. When new entities are created, they are sent from the "Created Exit" of the "Create Step" and a "Transfer Step" step can be used to transfer them from freespace (where they are created) to a particular node - in this case, they can be transferred either to the

Output@Server1, where they can exit with the parents, or they can be transferred into a "sink" where they can be counted.

In Simio simulation, for generating offsprings, a "children" entity object (similar to MP and FP) are added that are animated in such a manner that one can see the difference between children and parents. This is not required, but may be desirable at times. To create offsprings, within the "Processes" window, use the "Create Step" to change the "Object Instance Name" to "Children" and "Number of Objects" to "Random.Discrete" distribution based on information provided regarding how many children per couple to create. For example, Figure 7 shows that 10% of the population has 0 children, 20% has 1 child, 30% has 2 children, 30% has 3 children, 8% has 4 children and 2% have 5.

To graphically depict the children leaving the Server1 (population growth) with the parent (instead of them all leaving simultaneously on top of each other), one can change the "Path Allow Passing" property to "False" so they are shown in a line (Figure 8). Data collected from simulations can provide information regarding the number of off springs calculated. The parents are never split, as a separator object is not used – but stay together with the animated parent entity.

This population growth submodel is further used in an expanded consanguinity model by splitting it into two submodels: consanguineous marriages and non-consanguineous marriages. In the same way as in the population growth model, the male (MP) and female (FP) population is used followed by the separation of this population into two groups based on sex distribution:

male (59.5%) and female (40.5%). This sex distribution data is taken from World Factbook for Saudi Arabia [49]. First group have male (MP_C) and female (FP_C) population for consanguineous marriages. First group has male (MP_C) and female (FP_C) populations for consanguineous marriages, distribution rate for consanguineous union is 30%. We used to assign particular weights to paths i.e. for MP_C and FP_C we assign 35.7. Second group hold male (MP_NC) and female (FP_NC) population for non-consanguineous marriages and their paths weights were 65.9 (MP_NC) and 64.2 (FP_NC). These weights demonstrate the distribution of these population entities. Then we use combiner object to attach both male and female entities together. This distribution is based on average rates of marriages between first cousins among Saudi populations [50]. After reaching the queue of server named "Population Growth", birth events will occur and generate offsprings randomly. The same process takes place with non-consanguineous marriages and birth events. The population is continuously updated based on marriages and generated offsprings (Figure 10).

## 5. Results and Discussion

We generate a comprehensive report to provide particular statistics regarding the consanguinity model. To report results, we run simulations of the consanguinity model for at least 10 runs and results were exported in an Excel sheet. Here, we will discuss results with the help of tables generated through Simio pivot grid or reports. Our results contain four main categories which are: objects name, data source, category and value. Object name indicates the names of objects we used in our model which help to interpret values of objects. Data source shows the quality of data and category identifier used for high-level categorizing of statistics in reports (e.g. throughput). Value shows the quantity of entities passing through different objects.

Table. 1 demonstrates the total number of population entities created during the simulation runs as depicted in value column. Table 2 shows that the entities that enter are processed and exit the server object. The entities that individually pass through the paths which are 14 in numbers are also shown in Table 2. Table 3 shows the rest of object statistics such as total number of female and male population, consanguineous and non-consanguineous marriages, population growth and new population which is in process while passing through the server objects. Any model can be validating by importing and exporting data in Excel sheet through this SIMIO software.

## 6. Model Validation

As our main contribution is to provide a DEVS framework to model consanguinity and then demonstrate how a consanguinity model can be simulated in modern simulation software Simio, but the presented simulation model of consanguinity can easily be validated across the data available in different research studies. The data available in these studies are mostly presented in tables such as in 1974 a study was presented which provides quite uselful statistics regarding consanguinity[51], likewise M. Afzal*, et al.* has also provided a statistical based survey to access the "prevalence of consanguinous marriages, and the differentials by age at marriages,fertility and mortality experiences of the women who were married to their cousins and others"[52]. Another important study which can be use to validate our model was presented in 2001 by Rittler, M., et al. which was based on the statistical data obtained from Latin-American

Collaborative Study of Congenital Malformations (ECLAMC) during the period from 1967 to 1997 to analyzed the association between parental consanguinity and congenital anomalies [53]. Similarly many other studies conducted in Saudia [54] and Norway [55] can also be useful in vatidating our model.

## 7. Conclusions and Future Work:

In this study, we have developed a DEVS model of a population and used to study the *in silico* emergence of consanguinity in the offsprings. The main idea of this study was to take the first steps towards answering questions such as: what are the rates of consanguinity which can cause an increasing impact on the emergence of birth defects in a population? Therefore, it is important to model consanguinity in order to acquire a complete picture of congenital defects trends. Our contribution is a DEVS-based model of consanguinity which reveals a new population with a probability of congenital disorders due to consanguineous and non-consanguineous marriages followed by simulation using the Simio simulation software. Our results show that DEVS can be used effectively to model biological problems. In the future, we plan on applying the DEVS formalism to specific congenital disorders due to consanguinity.

## Acknowledgments

The authors wish to acknowledge the use of Simio simulation software Services. This work was made possible by an academic grant of Simio LLC to COMSATS Institute of IT.

**Table 1: Statistics of total number of consanguinity model entities created.** Generated results contain main categories which are: objects name, data source and value. This table demonstrates the total number of population entities created during the simulation runs. In the value column, we can observe the quantity of entities passing through different objects. Here, we notice that the total number of offsprings generated via consanguineous (Child_C) and non-consanguineous (Child_NC) marriage server object was 225 and 245 while the total number of female population (FP) and male population (MP) was 845 and 233 respectively. Here, the object name column indicates the names of the objects used in the model. While the data source column shows the properties of the objects used in the model e.g. Here, all objects have the "dynamic object" property which means that all of the objects generate dynamic entities. This assists in quickly viewing the total numbers of generated entities or individuals.

| Objects Name | Data Source | Value |
|---|---|---|
| Child_consanguineous | [Dynamic Object] | 225 |
| Child_non-consanguineous | [Dynamic Object] | 245 |

| Female population | [Dynamic Object] | 845 |
| Male population | [Dynamic Object] | 233 |

**Table 2: Statistics of consanguineous and non-consanguineous marriages including path entities.** This table shows the total number of consanguineous and non-consanguineous marriages. Total numbers of candidate couples [ParentInputBuffer] for consanguineous marriages were 91 while couples selected for consanguineous marriage that enter (MemberinputBuffer), processed and exit (OutputBuffer) the consanguineous marriage server object were 74 respectively. Total numbers of couples selected for non-consanguineous marriage that enter (MemberinputBuffer), processed and exit (OutputBuffer) the non-consanguineous marriage server object were 152 while total numbers of candidates were 158. The total path objects used in this model were 14 which show the entities (individuals) passes through these paths**.**

| Objects Name | Data Source | Value |
| --- | --- | --- |
| Consanguineous marriage(s) | [MemberinputBuffer] | 74 |
| Consanguineous marriage(s) | [OutputBuffer] | 74 |
| Consanguineous marriage(s) | [ParentInputBuffer] | 91 |
| Consanguineous marriage(s) | [Processed] | 74 |
| Non-consanguineous marriage(s) | [MemberinputBuffer] | 152 |

| Non-consanguineous marriage(s) | [OutputBuffer] | 152 |
| Non-consanguineous marriage(s) | [ParentInputBuffer] | 158 |
| Non-consanguineous marriage(s) | [Processed] | 152 |
| Path1, Path2, Path3, Path4, Path5, Path6, Path7 | [Travelers] | 92,161,78,155, 92,75,158 |
| Path8, Path9, Path10, Path11, Path12, Path13, Path14 | [Travelers] | 153,74,152,588, 596,225,245 |

**Table 3: Consanguinity Model Entities Statistics.** This table shows the object statistics such as total population growth as a result of consanguineous marriages that enter (inputBuffer), processed and exit (OutputBuffer) the population growth-consanguineous marriage server object, total population growth as a result of consanguineous marriages that enter (inputBuffer), processed and exit (OutputBuffer) the population growth_non-consanguineous marriage server object. Given table also demonstrates new population as a result of consanguineous and non-consanguineous marriages which is in process while passing through the server objects.

| Objects Name | Data Source | Value |
|---|---|---|
| Population growth_Consanguineous marriage(s) | [InputBuffer] | 299 |
| Population growth_Consanguineous marriage(s) | [OutputBuffer] | 294 |
| Population growth_Consanguineous marriage(s) | [Processed] | 295 |
| Population growth_Non-consanguineous marriage(s) | [InputBuffer] | 395 |
| Population growth_Non-consanguineous marriage(s) | [OutputBuffer] | 298 |

| Population growth_Non-consanguineous marriage(s) | [Processed] | 299 |
| NewPopulation_ Consanguineous marriage(s) | [InputBuffer] | 586 |
| NewPopulation_ Non-Consanguineous marriage(s) | [InputBuffer] | 594 |

**Figure captions**

**Figure 1: Degree relationships between family members**. Each circle represents the degree of consanguinity (genetic relatedness) with the person in the middle of the figure. (Figure adapted from [35] ).

**Figure 2: The basic entities in modeling and simulation and their relationship to each other.** Experimental frame specifies the conditions or environment in which system is experimented with, source system is the real or virtual system which is to be modeled, model is a mathematical representation of any system or structure and data is usually gathered by observing it and simulator is a software which follows model instructions and generate particular behavior of a real system.

**Figure 3: Basic entities of modeling and simulation of consanguinity.** This figure shows the main entities of consanguinity model which specifies experimental frame (Population, Allelic frequency, region, religion, cousin and consanguinity type, etc.), source system (consanguinity), model (DEVS -based) and simulator (Simio simulation software) for DEVS-based model of consanguinity.

**Figure 4: Flow diagram of population growth.** This flow diagram shows population growth model which initialized with two populations i.e., male (MP) and female (FP). After separating both populations, marriage events occur followed by population growth. Under population growth, another source of offspring (child) is added to generate children from resulting marriages randomly. This leads to a new population consisting of the number of total marriages and offspring.

**Figure 5: Flow diagram of consanguinity as a risk factor.** MP, male population; FP, female population; Child_C, children resulted from consanguineous union; Child_NC, children resulted from non-consanguineous union. The flow diagram shows DEVS-based consanguinity model. In this model, initially the male (MP) and female (FP) is used. After combining both male and female populations, the group was split into two, i.e. consanguineous and non-consanguineous marriages. Consanguineous and non-consanguineous marriages event now occurred separately followed by population growth (births) which include offspring source to generate children randomly. At the end, new population with the probability of consanguineous and non-consanguineous population with their rate of offspring growths is created.

**Figure 6: A 2D view of Simio complete population growth submodel.** This figure shows the DEVS-based population growth submodel, using Simio simulation software. To initialize this model, two source objects are used to create the male (MP) and female (FP) populations. Then both of the source populations are combined to a 'Combiner' object that takes one of each entity, combines them (marriage), and sends them to Server1 named "Population Growth". Within this process, new entities (offspring) are created. When new entities are created, they can be transferred into a "sink" (New Population) where they can be counted.

**Figure 7: Snapshot of "Create Step" for random offspring generation.** This figure shows the Properties for creating offspring randomly. To create offsprings, within the "Processes" window, use the "Create Step" to change the "Object Instance Name" to "Child" and "Number of Objects" to "Random.Discrete" distribution based on information provided regarding how many children per couple to create.  For example, this figure shows that 10% of the population has 0 children, 20% has 1 child, 30% has 2 children, 30% has 3 children, 8% has 4 children and 2% have 5.

**Figure 8: Parent and offspring move together into a new population object as discrete entities in a line.** This figure shows the children leaving the Server1 (population growth) with the parent graphically and then counted in sink object (new population).

**Figure 9:  A 3D view of the Simio complete population growth submodel.** This figure shows the DEVS-based population growth submodel in 3D, using Simio simulation software.

**Figure 10: Model of Consanguinity.** MP_C, male population for consanguineous marriages; FP_C, female population for consanguineous marriages; Child_C, children resulted from consanguineous union; Marriage_C, consanguineous union; PopulationG_C, consanguineous population growth; NewPopulation_C, new consanguineous population; MP_NC, male population for non-consanguineous marriages; FP_NC, female population for non-consanguineous marriages; Child_NC, children resulted from non-consanguineous union; Marriage_NC, non-consanguineous union; PopulationG_C, non- consanguineous population growth; NewPopulation_C, new non-consanguineous population.

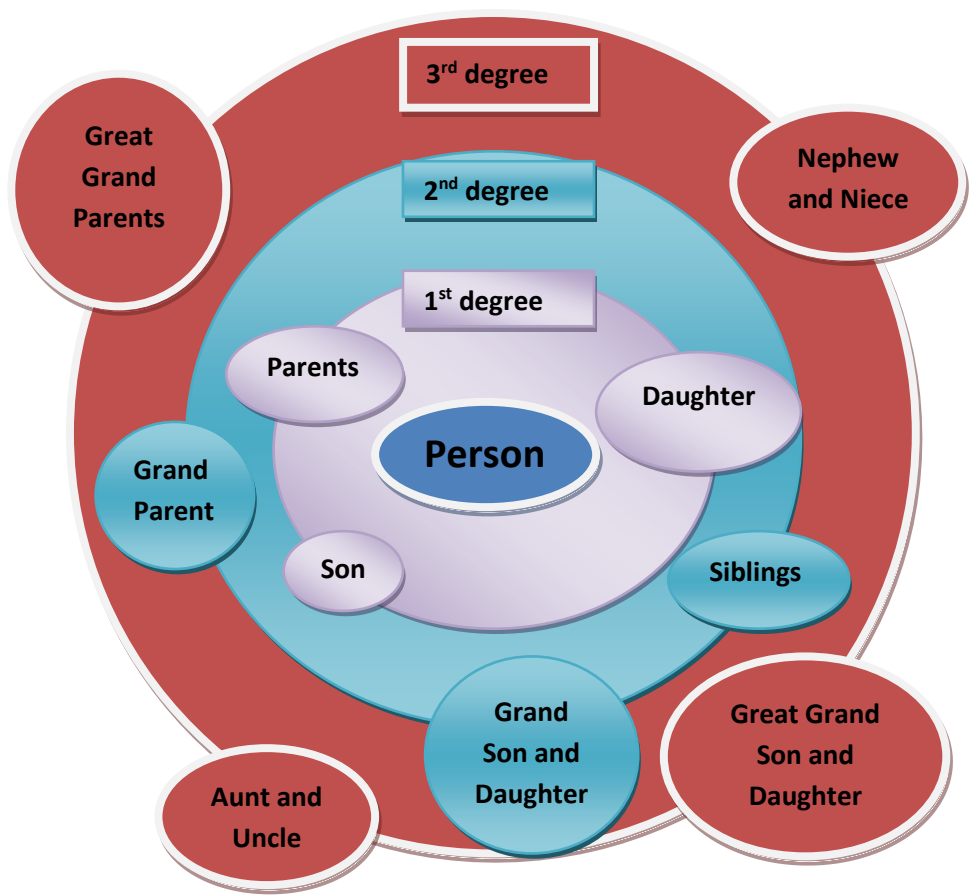

**Figure 1.**

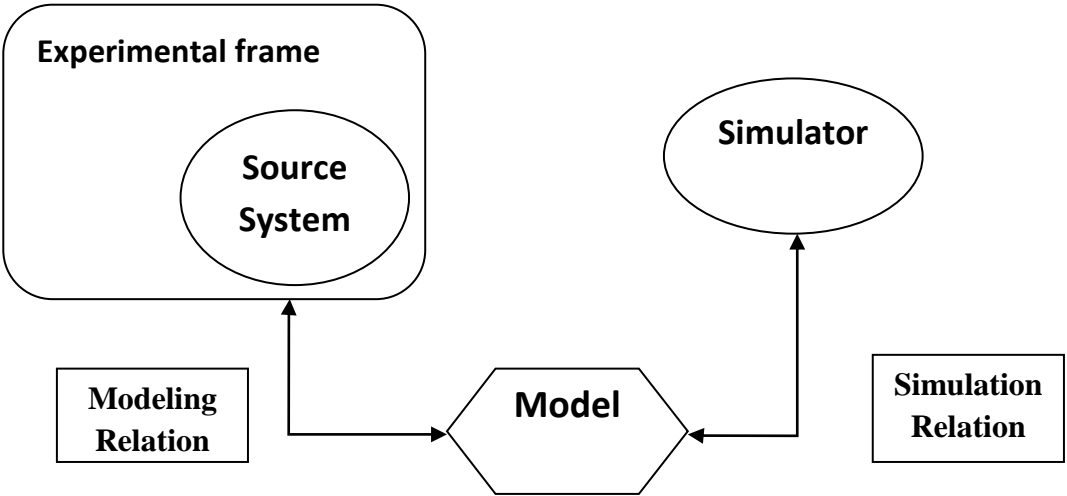

**Figure 2.**

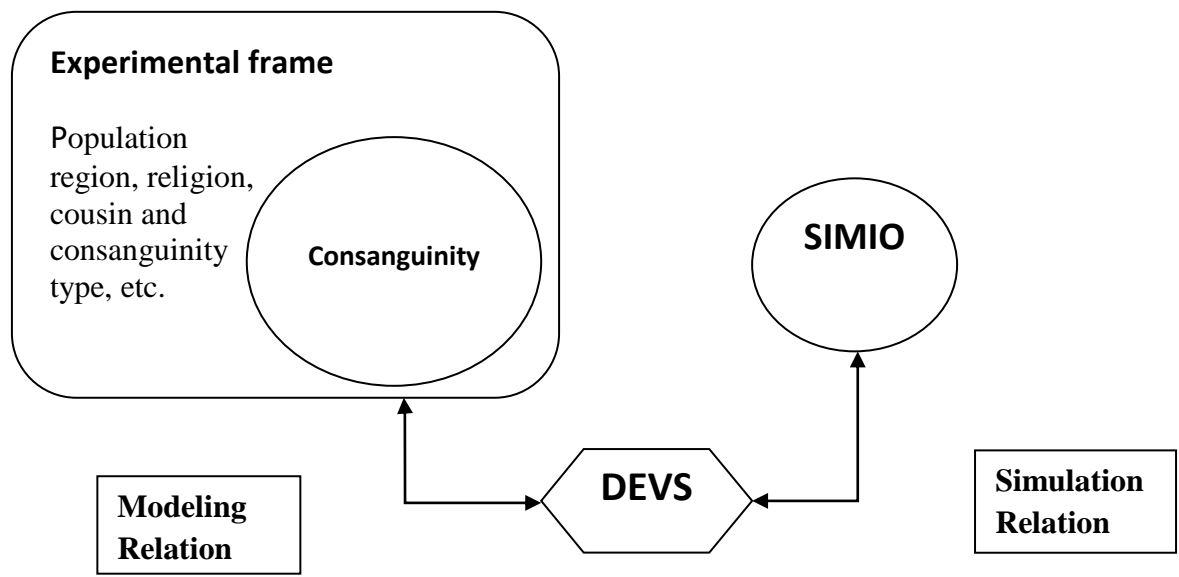

**Figure 3.**

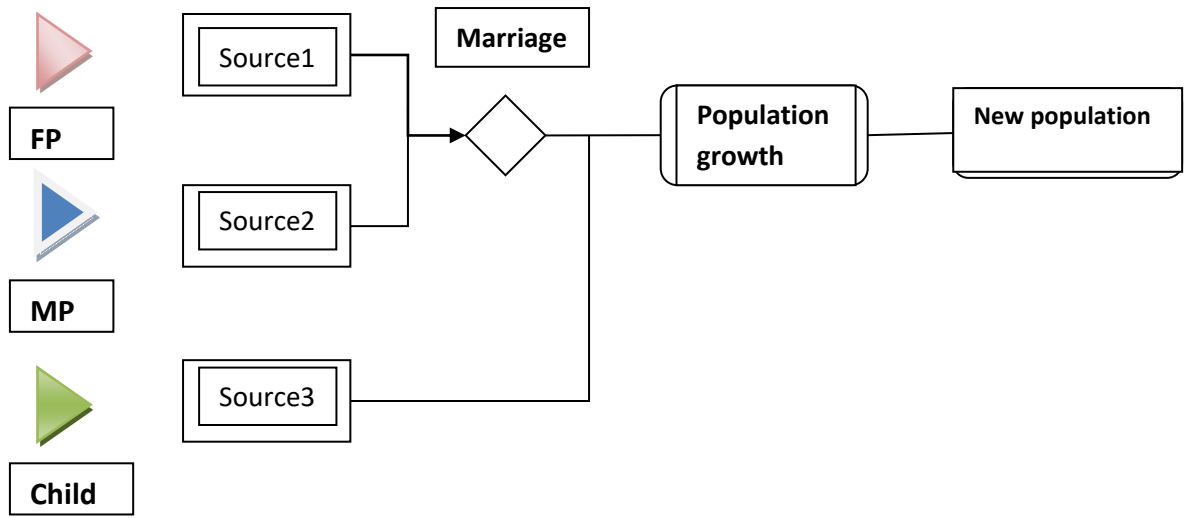

**Figure 4.**

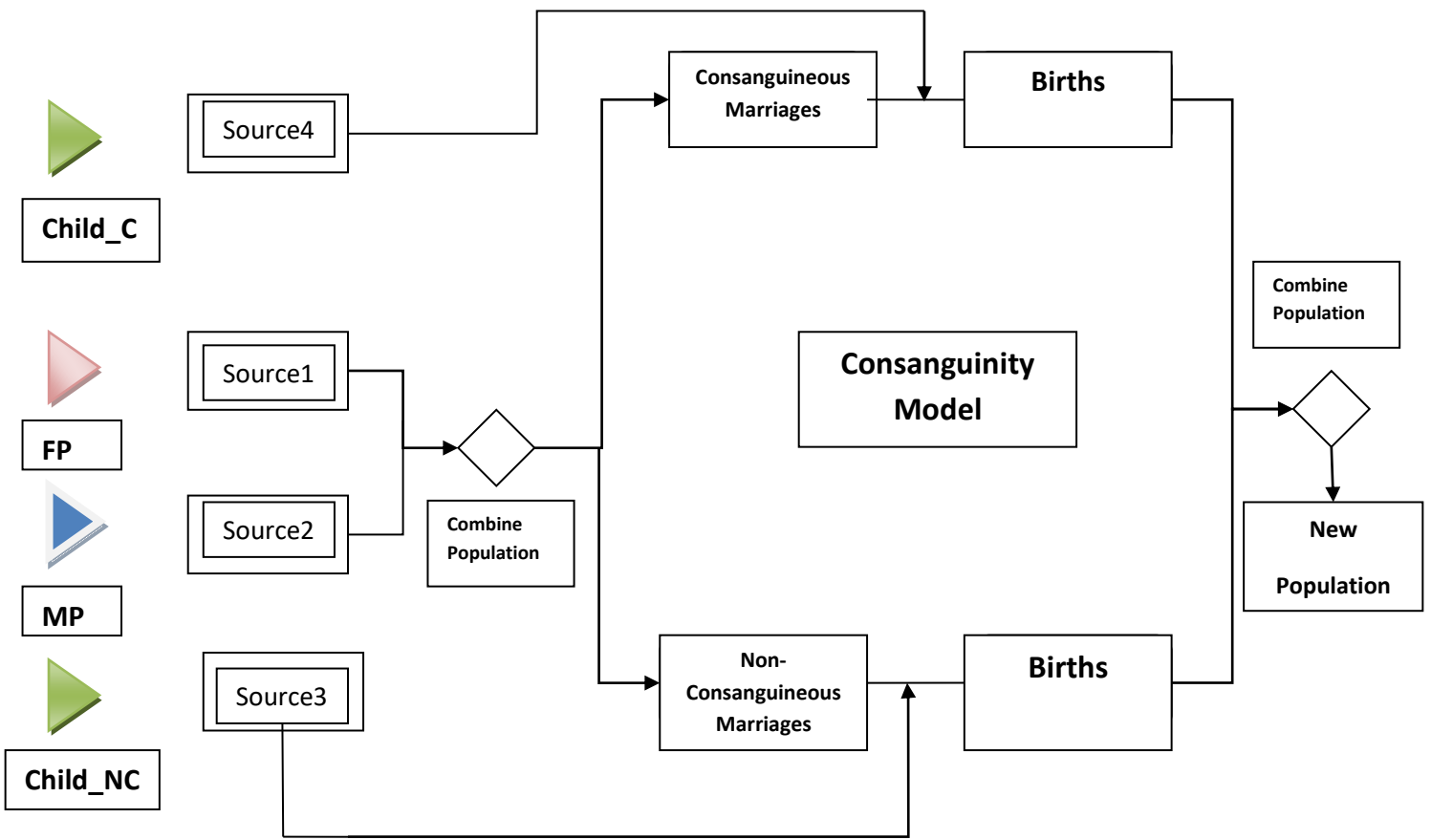

**Figure 5.**

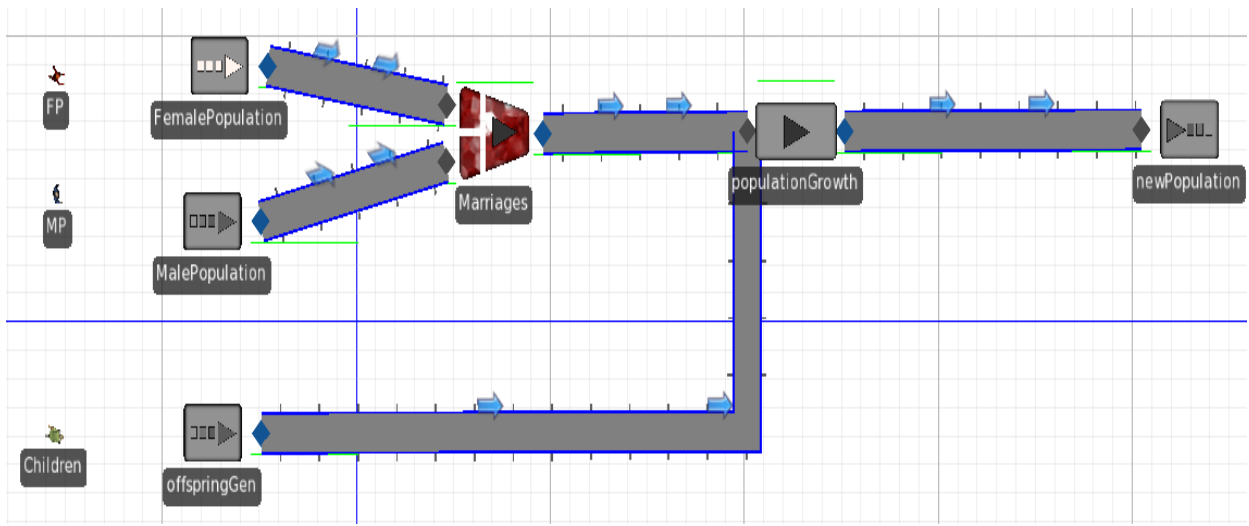

**Figure 6.**

**Figure 7.**

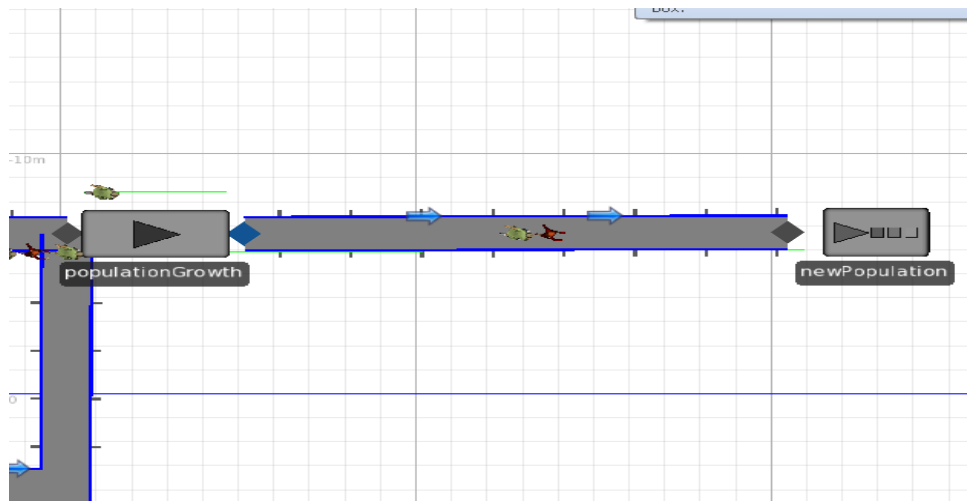

**Figure 8.**

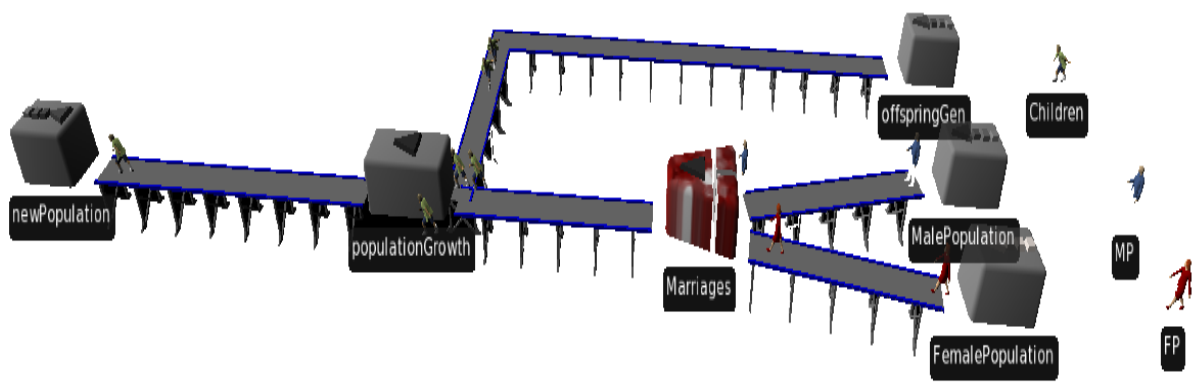

**Figure 9.**

**Figure 10.**